\documentclass[preprint, colorlinks, linkcolor=refcol, citecolor=refcol, urlcolor=refcol, a4paper, fleqn]{cas-dc}
\usepackage{xcolor}
\definecolor{refcol}{RGB}{178,34,34}
\usepackage{placeins}
\usepackage{physics}
\usepackage{amssymb}
\usepackage{amsmath}
\usepackage{amsthm}
\usepackage{orcidlink}    
\usepackage{dsfont}
\usepackage[numbers,sort&compress]{natbib}
\def\tsc#1{\csdef{#1}{\textsc{\lowercase{#1}}\xspace}}
\tsc{WGM}
\tsc{QE} 
\definecolor{copygreen}{HTML}{06402B}
 
\definecolor{irdatblue}{HTML}{328da8}

\begin{document}
\let\WriteBookmarks\relax
\def\floatpagepagefraction{1}
\def\textpagefraction{.001}

\shorttitle{Finite size effects on the phase diagram and the baryon fluctuations via momentum space constraints}
\shortauthors{Gy.~Kovács}

\title [mode = title]{Finite size effects on the phase diagram and the\\baryon fluctuations via momentum space constraints}
\tnotemark[T1]
\tnotetext[T1]{Contribution to EMMI Workshop: Aspects of Criticality II}

\author[L1,L2]{\color{black} {Gy}őző Kovács}[orcid=0000-0003-3735-7620]
\ead{gyozo.kovacs@uwr.edu.pl}
\affiliation[L1]{organization={Wigner RCP},
            addressline={Konkoly-Thege Mikos ut 29},
            city={Budapest},
            postcode={1121},
            country={Hungary}}
\affiliation[L2]{organization={University of Wrocław},
            addressline={Plac Maxa Borna},
            city={Wrocław},
            postcode={50204},
            country={Poland}}


\begin{abstract}
The effect of the finite system size on the QCD phase diagram was studied with various momentum space constraints within a mean-field quark-meson model. On the one hand side, the choice of the scenario -- low-momentum cutoff and discretization with periodic or antiperiodic boundary conditions -- and the presence of the vacuum fluctuations were found to strongly affect the volume dependence of the CEP. On the other hand, its location is significantly shifted in each case for linear sizes below $L\approx 10$ fm. This is also reflected in the conserved charge fluctuations, which were investigated along the phase boundary using multiple scenarios for the finite size effects. 
\end{abstract}




\maketitle

\hypersetup{
pdftitle={Finite size effects on the phase diagram and the baryon fluctuations via momentum space constraints},
pdfsubject={Contribution to EMMI Workshop: Aspects of Criticality II},
pdfauthor={Gy. Kovacs}
}

\section{Introduction}
\label{sec:Intro}
Although the phase diagram of strongly interacting matter has been already investigated extensively, it remained an open question whether there is a second-order critical endpoint (CEP) that would separate the already understood crossover from the expected first-order transition at high chemical potentials. The aim of finding the CEP is one of the main motivations of many recent works on both the theoretical and the experimental sides. One of the important differences between these sides of the search is the system size, which is either finite or infinitely large. The fireball formed in a heavy-ion collision -- that is expected to cross the phase transition in experiments -- always has a finite spatial extent. In contrast, the effective model calculations -- mostly used in theory to explore the phase diagram at finite chemical potential -- are solved in the thermodynamic limit, where the volume tends to infinity. To account for the finite size effects, it is usual to implement some kind of momentum space constraint in the theoretical models \cite{Palhares:2009tf, Fraga:2010qef, Magdy:2015eda, Magdy:2019frj, Braun:2004yk, Braun:2005gy, Braun:2005fj, Tripolt:2013zfa, Almasi:2016zqf, Klein:2017shl, Bhattacharyya:2012rp, Bhattacharyya:2014uxa, Pan:2016ecs, Wang:2018ovx, Wang:2018qyq, Xu:2019gia, Wan:2020vaj, MataCarrizal:2022gdr, Abreu:2015jya, Abreu:2016ihk, Abreu:2017lxf, Ishikawa:2018yey, Li:2017zny, Luecker:2009bs, Xu:2020loz, Bernhardt:2021iql, Bernhardt:2022mnx, Shaikh:2024ewz}. This can be a discretization with a specific (in most cases periodic or antiperiodic rarely with some other) boundary condition or, taking into account only the effect of the missing lowest modes, a simple low-momentum cutoff. The different scenarios in various models lead to generally similar, but in detail different finite-size dependence. In \cite{Kovacs:2023kcn, Kovacs:2023kbv} multiple types of momentum space constraints were implemented in a mean-field quark-meson model to understand the source of these differences. In the same publications, the baryon fluctuations were already also partially investigated, which we will continue to discuss in the present work.

The conserved charge fluctuations are one of the main tools in the search for the CEP, due to their strong dependence on the correlation length \cite{Koch:2008ia, Stephanov:2008qz}.
For instance, the fluctuations of the net baryon number are frequently used. These can be calculated directly in the effective field theoretical models, while they can also be accessed via the proton number fluctuations in experiments \cite{HADES:2020wpc, PhysRevLett.126.092301, STAR:2021fge},
although the latter may show different behavior \cite{Kitazawa:2012at, Koch:2023oez, Marczenko:2024nge}.\footnote{Fruthermore, the conservation of the net baryon or proton number is manifested in a different way in current theoretical models (with grand canonical ensemble) and experiments (with canonical picture and exact conservation of particle numbers).}
As discussed, the size of the physical system is also different in field theoretical calculations and in experiments. Therefore, the conserved charge fluctuations are usually characterized by the cumulant ratios that are free from explicit volume dependence. However, these ratios can still implicitly depend on the system size \cite{Skokov:2012ds}. The size dependence of the cumulant ratios has been studied in effective approaches at vanishing \cite{Magdy:2019frj, Bhattacharyya:2014uxa} or low chemical potential \cite{Almasi:2016zqf} and near the CEP \cite{Bernhardt:2022mnx, Kovacs:2023kbv} as a function of the temperature or the reduced parameters $T/T^\text{CEP}$ and  $\mu_q/\mu_q^\text{CEP}$. Here we will also study the finite size effects on the baryon fluctuations along the phase boundary as a function of the chemical potential.

The paper is organized as follows. In Sec.~\ref{sec:Model} we briefly summarize how the different momentum space constraints can be implemented in the well-known mean-field quark-meson model. In Sec.~\ref{sec:Phasediag} a comparison of the finite size effects in the different scenarios is given, with special attention to the path and the existence of the CEP(s) and the chirally broken phase. Finally, in Sec.~\ref{sec:Baryon} we show, how the baryon fluctuations are modified, before giving our conclusion and outlook in Sec.~\ref{sec:Conclusion}.

\section{Quark-meson model at finite volume}
\label{sec:Model}

In this paper, we will use the quark-meson model in the mean-field approximation, which can be written at different levels of complexity. One can start with only two \cite{Schaefer:2007pw, Palhares:2009tf} or two plus one flavors \cite{Schaefer:2008hk, Magdy:2015eda}. The mesonic sector may contain only scalar and pseudoscalar fields or might be also extended with further degrees of freedom, e.g. vector and axial vector mesons \cite{Kovacs:2016juc, Kovacs:2021kas, Giacosa:2024epf} (which is called the ePQM model). One can also consider the statistical confinement by including the Polyakov loop variables. However, contrary to the treatment of the vacuum term \cite{Kovacs:2023kcn}, it turns out that these differences hardly modify the qualitative aspects of the finite size dependence \cite{Kovacs:2023kbv}. In this section, we only summarize the implementation of different momentum space constraints in the quark-meson model in general. The technical details of the most advanced version we have used, the ePQM model, can be found in \cite{Kovacs:2023kbv}.

\subsection{Momentum space constraints}
\label{sec:Contraints}

The simplest scenario we use is the implementation of a low-momentum cutoff. In this case, the momentum integrals are simply modified by a Heaviside function
\begin{align}\label{eq:lowcutted_integral}
\int\frac{d^3p}{(2\pi)^3} \rightarrow \int \frac{d^3p}{(2\pi)^3} \theta(p-\lambda) ,
\end{align}
which is directly applicable to both the fermionic vacuum and thermal fluctuations. In the former case, the renormalization ensures that there is no conflict between the $\lambda\equiv \pi/L$ lower cutoff and the regularization.

Despite the unintuitive meaning in direct space, the use of a lower cutoff is well motivated by the expected importance of the lowest modes for the phase transition. Further support for this scenario is provided by the results of model calculations with ideal boson gas \cite{Redlich:2016vvb}, where the effect of the low-momentum cutoff turned out to be very similar to the effect of the directly implemented finite spatial extent.

The scenario of discretization in momentum space follows naturally from the Fourier transformation if the system has a finite size in direct space. The shape of the finite system and the boundary condition imposed on its surface determine the momentum modes. In most cases, a cubic volume with equal sides of length $L$ and the simplest periodic boundary condition (PBC) or antiperiodic boundary condition (APBC) is assumed. In these cases the momentum integral is substituted with a sum that runs over the modes $p_i = 2n_i \pi/L = n_i\Delta p$ or $p_i = (2n_i+1)\pi/L = (n_i+1/2)\Delta p$, respectively, with $n_i \in \mathds{Z}, i \in (x,y,z)$, while $\Delta p\equiv 2 \pi /L$ is the size of the momentum grid. Note that the main difference between PBC and APBC is the presence of the zero mode, which is expected to have an important role in the case of the finite size effects. To ease the numerics it is useful to rearrange the summation of the momentum modes from a cubic pattern, where each point has to be counted separately, to a spherical summation (with $K$ being the corresponding kernel)
\begin{align} \label{eq:sum_cubetosphere}
\sum_{n_x,n_y,n_z=-\infty}^{\infty} K(p_i) = \sum_{j=1}^{\infty} \sum_m K(p_{j},m) ,
\end{align}
which makes use of the multiplicity $m$ of the modes with certain $\left|\vec{p}\right|$. Furthermore, it can be shown that the difference of the neighboring momenta $\Delta p_{j,j+1}=p_{j+1}-p_j$ decreases with the increasing $j$. The fermionic matter part is regulated by the Fermi-Dirac statistics, hence its integrand $K^\text{mat}(p_j,m)$ and also the change of this kernel from $j$ to $j+1$ is suppressed for large $j$. Due to the suppression, one can simply neglect the high momentum modes above a certain $\lambda_\Sigma^\text{cut}$ cutoff. Alternatively, one can change the summation to integration at an even lower $\lambda_\Sigma$, which is called the UV improvement \cite{Xu:2020loz, Bernhardt:2021iql}. Then, instead of the infinite summation one obtains
\begin{align} \label{eq:UVimproved_sum}
\frac{1}{L^3}\sum_{j=1}^{j_\text{max}}\sum_m + \int \frac{d^3p}{(2\pi)^3} \theta(p-\lambda_\Sigma) .
\end{align} 
In either case, $\lambda_\Sigma^\text{cut}$ or $\lambda_\Sigma$ has to be set separately for each size, such that the effect of the cutoff or the replacement is numerically negligible. 

The UV-divergent fermionic vacuum fluctuations might also be regularized with a higher cutoff. Moreover, Implementing the UV improvement and hence a spherical integration for high momentum allows the use of the usual renormalization. However, the ambiguity of fitting a spherical integral to a cubic grid would lead to a large oscillating error due to the too strong increase of the integration kernel. To apply the UV improvement in this case, one has to integrate from a cubic lower boundary. This raises only a technical difficulty, which is discussed in detail in the appendix of \cite{Kovacs:2023kbv}. In the end, instead of the summation one arrives with 
\begin{align} \label{eq:UVimproved_sum_cubic}
\frac{1}{L^3}\sum_{j=1}^{j_\text{max}}\sum_m  
&+ \int_{p=\lambda_\Sigma}^{p=\sqrt{3}\lambda_{\Sigma}} \frac{dp~p^2}{(2\pi)^3}\Omega^{\lambda_\Sigma}(p) \nonumber \\
&+\int \frac{d^3p}{(2\pi)^3} \theta(p-\sqrt{3} \lambda_\Sigma),
\end{align} 
where $\Omega^{\lambda_\Sigma}(p)$ is the solid angle corresponding to the points of a sphere of radius $p$ that are outside of the cubic box of side length $2\lambda_\Sigma$. In the radial momentum integration $\Omega^{\lambda_\Sigma}(p)$ replaces the full $4\pi$ solid angle in the given term. Finally, with the UV improvement, one can define a method to implement the discretization in the quark-meson model in both the fermionic vacuum and matter fluctuations, which allows renormalization and is also numerically applicable. We use this method in our calculations.

\section{Size dependence of the phase diagram}
\label{sec:Phasediag}

\subsubsection*{Low-momentum cutoff}

\begin{figure}[tb!]
    \centering
    \includegraphics[width=0.95\linewidth]{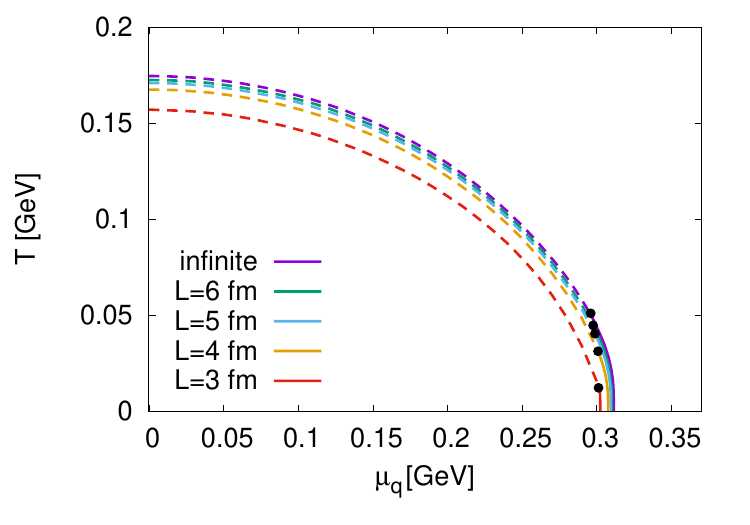}
    \includegraphics[width=0.95\linewidth]{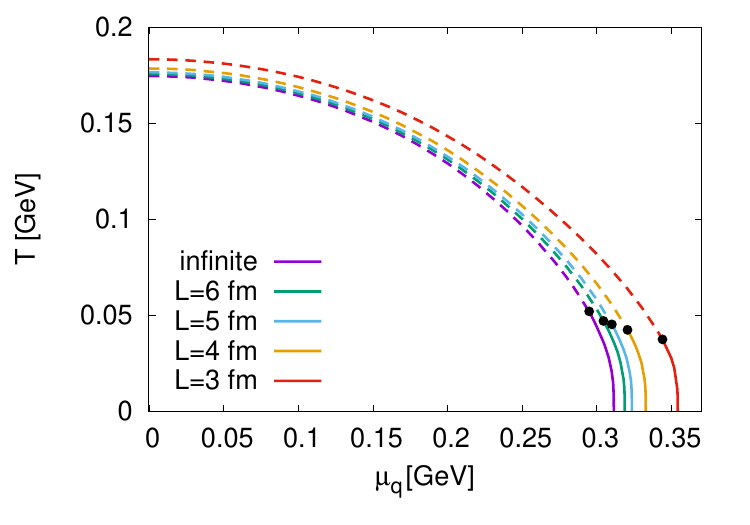}
    \caption{The size dependence of the phase diagram in the low-momentum cutoff scenario in the case modified (top) and unmodified (bottom) vacuum contribution.}
    \label{fig:PhasediagLowmom}
\end{figure}

To investigate the size dependence of the phase diagram,\footnote{We plot the phase diagram in the $T-\mu_q$ plane and have symmetric quark matter with $\mu_u=\mu_d=\mu_s\equiv\mu_q=\mu_B/3$.} and in particular the CEP, we apply the scenarios discussed in Sec.~\ref{sec:Contraints}, starting with the simplest case of having a low-momentum cutoff.
In previous works on finite size effects within the framework of the quark-meson model at mean-field (see, e.g., \cite{Palhares:2009tf, Fraga:2010qef, Magdy:2015eda, Magdy:2019frj}), the fermionic vacuum fluctuations were not included. Consequently, the effect of the finite volume on the physical quantities at $T=\mu_q=0$ was absent. However, the modification of the vacuum physics alters the size dependence of the whole phase structure.\footnote{Generally, modification of the vacuum contribution strongly affects the phase diagram already at infinite volume.} This can be clearly seen in Fig.~\ref{fig:PhasediagLowmom}, where the phase diagram is shown at infinite and finite sizes for the low-momentum cutoff scenario.
When the lower cutoff is applied for the vacuum contribution as well (top panel), the CEP and, at a somewhat smaller size, the whole chirally broken phase disappears (at $L\approx 2.5$ fm and $L\approx 2$ fm, respectively) contrary to the case of unmodified fermionic vacuum fluctuations (bottom panel). This gives rise to the difference observed between previous calculations in QM and other frameworks, namely the increasing instead of decreasing trend of the chirally broken phase with the decreasing system size. 

\subsubsection*{Momentum discretization}

When the momentum discretization is applied to the vacuum contribution the chiral symmetry breaking is even enhanced for decreasing system size for PBC and -- for not very small, i.e. $L>1$ fm sizes -- also for APBC unlike in the case of the low-momentum cutoff. For PBC this could be explained by the presence of the zero mode, while for APBC it shows that for these intermediate sizes, not only the zero but also further low modes contribute significantly. Comparing our results with mean-field calculations in the NJL model \cite{Bhattacharyya:2012rp, Bhattacharyya:2014uxa, Pan:2016ecs, Wang:2018ovx, Wang:2018qyq, MataCarrizal:2022gdr} and with results in functional approaches \cite{Tripolt:2013zfa, Bernhardt:2021iql} one can elaborate on the difference between the effect of discretization and low-momentum cutoff on the vacuum fluctuations. It seems that for PBC the difference comes from the mean-field treatment of the momentum integrals (the trend is similar in NJL but different in functional methods), while for APBC it comes from the treatment of the divergent vacuum term (the increase for intermediate sizes does not appear in the NJL model). 

Due to the enhanced spontaneous chiral symmetry breaking, the vacuum value of the chiral condensate(s) $\bar\phi$ increases with the decreasing system size. However, the grand potential is not bounded from below for $\bar\phi\to\infty$ if the vacuum term is present since it becomes dominant with a $\propto$\,$-\bar\phi^4 \log \bar \phi$ behavior. These together lead to the loss of a common solution for the field equations at a given finite size (e.g. for the ePQM around $L\sim 5.5$ fm). To avoid this problem, we focus on the cases where the vacuum contribution is not modified by the finite volume or the discretization is taken into account only in the lowest modes (as was done in \cite{Palhares:2009tf}) when studying the phase diagram. 

\begin{figure}[tb!]
    \centering
    \includegraphics[width=0.95\linewidth]{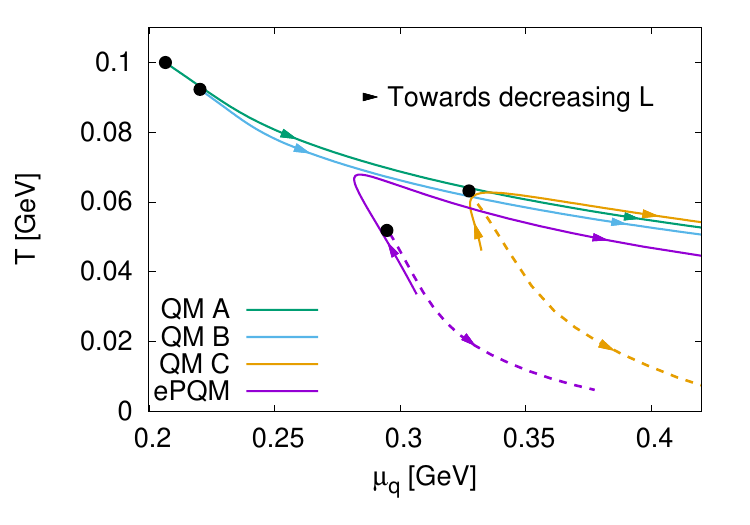}
    \includegraphics[width=0.95\linewidth]{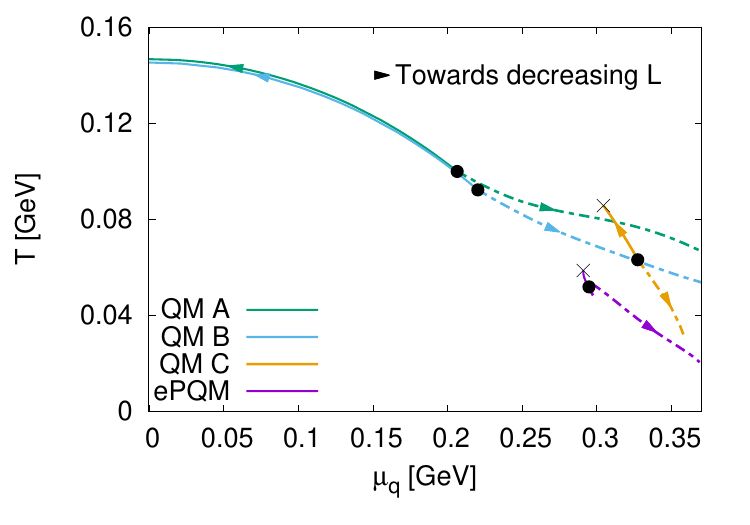}
    \caption{The size-dependent path of the CEP in different quark-meson models with antiperiodic (top) and periodic (bottom) boundary condition.}
    \label{fig:PhasediagDiscret}
\end{figure}

The discretization has an interesting effect also on the fermionic matter fluctuations. As the temperature goes to zero the Fermi-Dirac distribution tends to a unit step-function with the discontinuity at the Fermi surface at $p=\sqrt{\mu_q^2-m_f^2}$ for chemical potentials greater than the constituent quark mass $m_f$. When $\mu_q$ is increased at a fixed $T\gtrapprox 0$ this Fermi surface extends and passes the discrete modes at finite $L$ one by one. This leads to a sudden increase in the fermionic matter part and hence to a sudden drop in the chiral condensate(s), resulting in a "\textit{staircase-like}" multilevel phase transition. The different modes entering below the Fermi surface may generate separate first-order phase transitions and separate critical points. For sufficiently small sizes, the lowest mode always gives the leading CEP, but this is not necessarily the one that is connected to the infinite volume CEP. To better understand the resulting size dependence of the phase structure and especially the path of the CEPs, we investigate multiple quark-meson models. Following \cite{Kovacs:2023kbv}, we implement the simplest $N_f=2$ model introduced in \cite{Palhares:2009tf} (QM A), two parameterizations of the $N_f=2+1$ version in \cite{Schaefer:2008hk} (QM B and C) with $m_\sigma=600$ and $800$ MeV mass of the scalar meson,\footnote{The parameter sets for QM B and C can be found in the 11th and 13th row of the bottom panel of Table~II in \cite{Schaefer:2008hk}.} and the more advanced ePQM model. QM A-C has no vacuum fluctuations taken into account and the effect of the Polyakov loop is also absent contrary to the ePQM for which we use the setup and parameterization of Fit$^{1,1,1,2}$ in \cite{Kovacs:2016juc}. However, the path of the CEP in finite volumes is insensitive to these differences and is instead determined by its location at $L=\infty$. This can be seen in the top panel of Fig.~\ref{fig:PhasediagDiscret}, where the size dependence of the leading critical points is shown for APBC.
The solid lines correspond to the path of the critical points connected to the $L\to \infty$ CEP, while the dashed lines show the path of the critical points corresponding to leading CEP at very small sizes (when it is separated).
For models with a \textit{higher-lying} CEP at infinite size (QM A and B) the leading critical point has a smooth path from $L\to\infty$ to $L\to 0$, while for those with a \textit{lower-lying} CEP (QM C and ePQM), there is an interchange, which gives rise to a peculiar size dependence. Since this behavior is directly related to the structure of the momentum integrals it naturally appears also in NJL model calculations \cite{Xu:2019gia} as well as in quark-meson models with FRG \cite{Tripolt:2013zfa, Almasi:2016zqf}.
A further complication arises in the case of PBC, since then the CEP moves even to higher $T$ and lower $\mu_q$ for decreasing sizes if the vacuum fluctuations are not modified. Furthermore, since the zero mode in the matter part is not compensated by the zero mode in the vacuum part, below a certain size, the solution for high temperatures becomes absent. Including at least the effect of the zero mode also in the vacuum contribution solves this problem, while reverting the trend of the CEP to be similar to the APBC and the low-momentum cutoff scenario. The path of the CEPs for PBC is shown in the bottom panel of Fig.~\ref{fig:PhasediagDiscret}. Here the solid line corresponds to the case where the vacuum contribution is not modified, while the dashed-dotted line shows the case where the discretization is taken into account at least for the zero mode. In the latter case, for small sizes, the effect of having multiple critical points becomes important again, similar to what was observed for APBC.

\section{Baryon number fluctuations}
\label{sec:Baryon}

In this section, we discuss the finite size effects on the baryon number fluctuations. The generalized susceptibilities of the net baryon number, defined as
\begin{align} \label{eq:gensusc}
\chi_n =\left. \frac{\partial^n p/T^4}{\partial (\mu_q /T)^n}\right|_T,
\end{align}
are related to the cumulants via 
\begin{align}
C_n = V T^3 \chi_n.
\end{align}
Taking the ratios of the cumulants, which are equal to the corresponding susceptibility ratios, the explicit volume dependence can be canceled. For instance, one can define the skewness $S\sigma=C_3/C_2$ or the (excess) kurtosis $\kappa \sigma^2=C_4/C_2$, where $\sigma$ is the variance. However, these quantities may still have an implicit dependence on the system size \cite{Skokov:2012ds}.
In the mean-field calculations, the generalized susceptibilities can be easily obtained, since the grand potential is directly accessible and thus the pressure
\begin{align} 
p(T, \mu_q) \equiv \Omega (0,0) - \Omega (T, \mu_q)
\end{align}
can be determined in contrast to the case of the Dyson-Schwinger approach \cite{Bernhardt:2021iql, Bernhardt:2022mnx}. Here, we calculate the baryon fluctuations at finite system size both with discretization (using APBC) and with low-momentum cutoff.
In the discretization scenario, the volume dependence of the phase diagram is more complicated, especially in the vicinity of the CEP(s). The \textit{staircase-like} phase transition at low temperatures and high chemical potentials and the multiple critical points, discussed in Sec.~\ref{sec:Phasediag}, can strongly affect the conserved charge fluctuations already in the crossover regime \cite{Almasi:2016zqf}. 
Therefore, in the present case, we restrict our study to a model with a higher-lying CEP,\footnote{In the first-order region, one could still see the effect of the critical points corresponding to the higher modes entering below the Fermi surface. However, here we use only moderate system sizes, above $L=5$ fm, where this does not seem to affect the baryon fluctuations around $T=T_c$ and below $\mu_q=$ 250 MeV.} namely the parameterization with $m_\sigma=600$ MeV in \cite{Schaefer:2008hk}. This model has no fermionic vacuum fluctuations, and therefore this setup is equivalent to the vacuum contribution not being modified by the finite size effects. The corresponding effect is already clarified in Sec.~\ref{sec:Phasediag}, while this version of the model allows the implementation of both the low-momentum cutoff and the discretization for small sizes.
To properly recover the $\sigma^2\kappa = 1$ value of the kurtosis in the chirally broken phase, the Polyakov loop variables have to be also included for the statistical confinement. For this, we will use the usual implementation and the Polyakov-loop potential introduced in \cite{Lo:2013hla}. 
Consequently, the model we use is equivalent to the Polyakov loop extended version of QM B in Sec.~\ref{sec:Phasediag}.

\begin{figure}[tb!]
    \centering
    \includegraphics[width=0.9\linewidth]{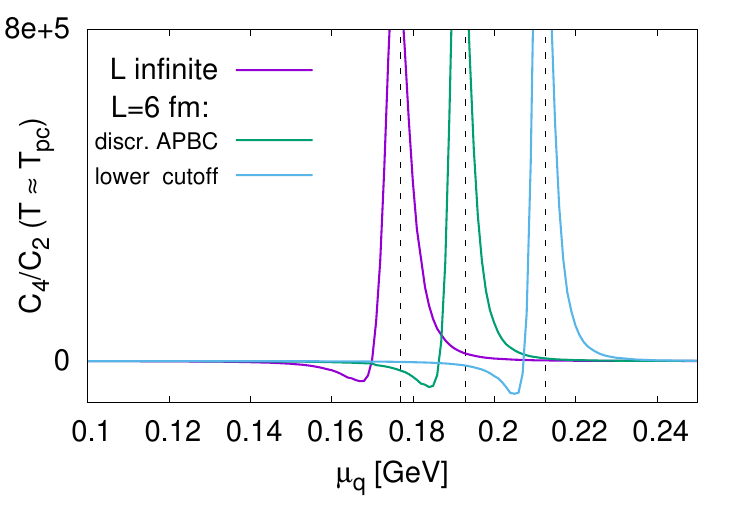}
    \includegraphics[width=0.9\linewidth]{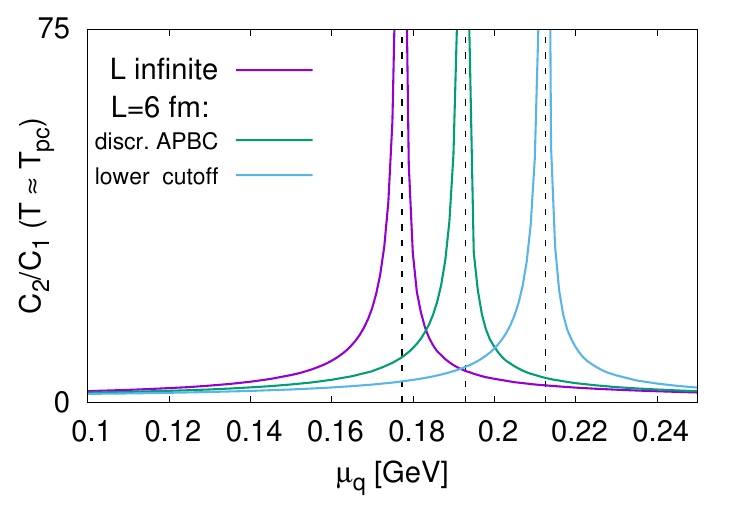}
    \includegraphics[width=0.9\linewidth]{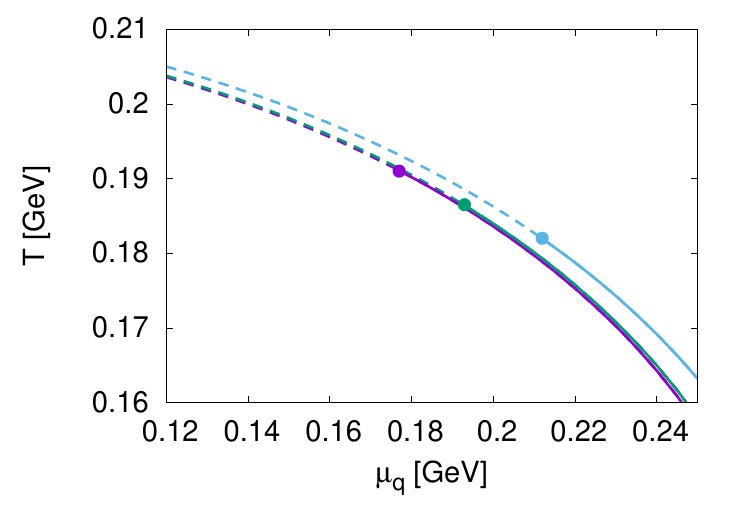}
    \caption{The ratio of the fourth and second (top) and the second and first (middle) order cumulants for infinite size and $L=6fm$ -- with discretization using APBC in green and with a low-momentum cutoff in blue -- along the respective phase boundary (bottom).}
    \label{fig:Kurtosis}
\end{figure}

In the framework described above, we calculate the ratio of the fourth and second (corresponding to the kurtosis) and the ratio of the second and first order cumulants as a function of the quark chemical potential at $T\lessapprox T_{pc}(\mu_q)$ (i.e. along but slightly below the phase transition line). These ratios are shown in the top and middle panel of Fig.~\ref{fig:Kurtosis},
while the bottom panel shows the corresponding phase boundaries.
In each plot, the purple curve belongs to $L=\infty$, while the green and blue curves are the results at $L=6$ fm with discretization using APBC and with low-momentum cutoff, respectively. It can be seen that the peak indicating the location of the CEP -- also shown on the bottom panel -- is shifted to higher chemical potentials for both scenarios. However, the change, and especially the extension of the chirally broken phase, is weaker in the case of discretization. The size of the signals around the CEP is only mildly modified since the divergences in still present and a real critical phenomenon can be seen when the momentum space constraints are applied to mean-field models.\footnote{Furthermore, the criticality is present even in a non-local model if the momentum dependence is treated with a form factor as in \cite{Sasaki:2006ww}.} To investigate the finite size remnants of the criticality one needs to implement different scenarios to account for the finite system size or other approximations. However, exploring these possibilities is beyond the scope of the present work.


\section{Conclusion}
\label{sec:Conclusion}

We investigated the QCD phase diagram and the baryon number fluctuations in a mean-field quark-meson model including the finite size effects via momentum space constraints. Several scenarios, including low-momentum cutoff and discretization with different boundary conditions, were implemented to show how they differ in the details at finite volumes. Despite the differences, it was found that the CEP is significantly shifted -- in most cases either to larger chemical potentials or to the lower temperatures -- below $L\approx 10$ fm. 

The finite size effects on the baryon fluctuations were also studied using low-momentum cutoff and discretization with APBC. It was found that the signal of the CEP is shifted, but its shape is hardly modified due to the unchanged criticality in the mean-field approaches when only the momentum space constraints are applied.

The results in the present work and in previous publications are either not complete (in the sense of removing the divergences at finite sizes) \cite{Kovacs:2023kbv} or have too low resolution \cite{Bernhardt:2022mnx} to see the scaling behavior near the CEP. To better understand the effect of the finite system size on the critical fluctuations in the QCD phase diagram -- which would be interesting for the recent and near-future measurements -- we need to improve the implementation of the finite size effects in the effective model calculations. 

\section{Acknowledgement}
We would like to thank P.M. Lo, K. Redlich, C. Sasaki, P. Kovács, and Gy. Wolf for the valuable discussion.
Gy.~K. acknowledge support by the Hungarian National Research, Development and Innovation Fund under Project number K 138277. The work of Gy.~K. is partially supported by the Polish National Science Centre (NCN) under OPUS Grant
No. 2022/45/B/ST2/01527.

\newpage
\appendix

\FloatBarrier

\bibliography{proc_EMMI_Wroclaw}

\end{document}